\documentclass[%
reprint,
amsmath,amssymb,
aps,
prb,
]{revtex4-2}
\usepackage{graphicx}
\usepackage{CJKutf8}
\usepackage{epstopdf}
\usepackage{subfigure}
\usepackage{pifont}
\usepackage{bm}
\usepackage{xcolor}
\usepackage[colorlinks=true,linkcolor=blue,anchorcolor=blue,citecolor=blue,urlcolor=blue]{hyperref}
\begin{document}
\begin{CJK}{UTF8}{gbsn}
\preprint{APS/123-QED}

\title{Spin excitations in bilayer La$_3$Ni$_2$O$_7$ superconductors with the interlayer pairing}

\author{Meiyu Lu$^1$}
\author{Tao Zhou$^{1,2}$}%
\email{tzhou@scnu.edu.cn}

\affiliation{$^1$Guangdong Basic Research Center of Excellence for Structure and Fundamental Interactions of Matter, Guangdong Provincial Key Laboratory of Quantum Engineering and Quantum Materials, School of Physics, South China Normal University, Guangzhou 510006, China\\
	$^2$Guangdong-Hong Kong Joint Laboratory of Quantum Matter, Frontier Research Institute for Physics, South China Normal University, Guangzhou 510006, China}

\date{\today}

\begin{abstract}

Prompted by the recent discovery of high-temperature superconductivity in La$_3$Ni$_2$O$_7$ under pressure, this study delves into a theoretical investigation of spin excitations within this intriguing material. 
Through self-consistent mean-field calculations, we propose that superconductivity in this compound
  is primarily driven by interlayer pairing mechanisms. This interlayer pairing results in an effective $s_\pm$
  pairing gap, with the sign of the pairing gap varying across different Fermi pockets.
In the superconducting state, our analysis reveals a striking absence of a spin resonance mode. Moreover, we reveal a spectrum of energy-dependent incommensurate spin excitations. The observed incommensurate structures are elegantly explained by the nesting effect of energy contours, providing a coherent and comprehensive account of experimental observations.
The implications of these spin excitations in La$_3$Ni$_2$O$_7$ are profound, offering critical insights into the superconducting mechanism at play. Our results not only contribute to the understanding of this novel superconductor but also pave the way for further research into the interplay between spin dynamics and unconventional superconductivity in layered materials.

\end{abstract}

\maketitle


\section{\label{sec1.intr}INTRODUCTION}

The recent discovery of bilayer nickelate superconductors, exemplified by La$_3$Ni$_2$O$_7$, represents a significant milestone in the field of high-temperature superconductivity~\cite{s41586-023-06408-7,Wang_2024}. With a superconducting transition temperature (T$_c$) of 80 K under pressure, this material has emerged as a novel addition to the family of high-T$_c$ superconductors. Intriguingly, the nickel ions in this compound exhibit an average valence of $+2.5$, corresponding to a $3d^{7.5}$ electronic configuration - a striking departure from the more commonly observed $3d^9$ configuration found in cuprate superconductors and earlier infinite-layer nickelate superconductors~\cite{RevModPhys.78.17,Li2019}.

Density-functional-theory calculations have elucidated the low-energy electronic structure of La$_3$Ni$_2$O$_7$, revealing that it is predominantly shaped by the Ni-$3d_{z^2}$ and Ni-$3d_{x^2-y^2}$ orbitals~\cite{PhysRevLett.131.126001}. This insight has led to the widespread adoption of the bilayer two-orbital model, which encompasses four energy bands, as a framework for understanding the normal state energy bands~\cite{PhysRevLett.131.126001}.

Over the past year, extensive theoretical and experimental research has been dedicated to the La$_3$Ni$_2$O$_7$ compound~\cite{Wang_2024}. Studies have shown that as pressure increases beyond 100 GPa, the evolution of the Ni-$3d_{z^2}$ electron density closely mirrors changes in the superconducting transition temperature~\cite{li2024pressure,huo2024modulation}. This correlation suggests that the Ni-$3d_{z^2}$ orbital plays a crucial role in the superconductivity of La$_3$Ni$_2$O$_7$. Unlike traditional layered superconductors, where pairing primarily occurs within individual layers, the strong interlayer hopping of $d_{z^2}$ orbitals in this compound indicates that interlayer interactions could significantly influence the pairing process. Consequently, the interplay between intra- and interlayer pairing has become a central focus of research~\cite{arXiv2306.07275,PhysRevB.108.165141,PhysRevLett.131.236002,PhysRevLett.132.106002,PhysRevB.108.L140504,lu2023superconductivity,lange2024pairing,schlomer2023superconductivity,PhysRevB.108.L201108,arXiv2310.02915,PhysRevB.108.174511,arXiv2311.05491,wu2024,PhysRevLett.132.146002,arXiv2402.07449,PhysRevB.110.L060506,yang2024decom,PhysRevLett.132.036502,fan2023superconductivity,arXiv2309.05726,arXiv2309.15095,PhysRevB.108.174501,arXiv2401.15097,arXiv2308.16564,PhysRevB.108.214522,arXiv2310.17465,arXiv2308.09698}.

A number of studies have highlighted the significance of interlayer pairing, which is driven by robust interlayer interactions and is further amplified by substantial Hund's coupling~\cite{PhysRevB.108.L201108,arXiv2310.02915,PhysRevB.108.174511,arXiv2311.05491,wu2024,PhysRevLett.132.146002,arXiv2402.07449,PhysRevB.110.L060506,yang2024decom}. It has been hypothesized that this interlayer pairing scenario could account for the reduced superconducting transition temperature observed in trilayer nickelate superconductors~\cite{PhysRevB.110.L060506}. Experimental data from neutron scattering and resonant inelastic X-ray scattering have corroborated that interlayer magnetic superexchange interactions are significantly stronger than intralayer ones~\cite{xie2024neutron,chen2024electronic}. While the La$_3$Ni$_2$O$_7$ material has attracted considerable attention for its unique pairing mechanism and symmetry, the intricate interplay between inter- and intralayer interactions is still an area ripe for further exploration. A more profound understanding of these interactions could illuminate the superconducting properties and may lead to advancements in enhancing the material's performance.

While the mechanism behind high-temperature superconductivity remains enigmatic, spin excitations are believed to play a pivotal role~\cite{Moriya_AdvPhys_2000_v49_p555}. The spin excitations in various unconventional superconducting systems have been the subject of intensive research. For La$_3$Ni$_2$O$_7$, experimental reports of magnetic order in the parent compound suggest that magnetic excitations may exist in the superconducting state and could be integral to superconductivity~\cite{xie2024neutron,chen2024electronic}. Over the past year, the study of normal state spin excitations has been extensive, with the pairing mechanism and symmetry being explored through the lens of spin fluctuations~\cite{PhysRevB.108.L140505,PhysRevLett.132.106002,arXiv2306.07275,PhysRevLett.131.236002,Zhang2024,PhysRevB.108.165141,PhysRevB.108.L201121}. Despite this, a consensus for the pairing mechanism and the pairing symmetry has yet to be reached.

Spin excitations in the superconducting state are known to be sensitive to the normal state Fermi surface and the superconducting pairing symmetry, making them valuable probes for the pairing symmetry of unconventional superconductors. Recently, the possibility of magnetic excitations and spin resonance has been proposed, considering both $d$-wave and typical $s_\pm$ pairing symmetries~\cite{PhysRevB.109.L180502}.

In this study, we delve into the superconducting pairing term at the mean-field level, examining the competition between interlayer and intralayer pairing channels. We hypothesize that superconductivity in La$_3$Ni$_2$O$_7$ arises from interlayer pairing within the Ni-$3d_{z^2}$ orbital. Additionally, we investigate the frequency and momentum dependencies of spin excitations in the superconducting state, aiming to provide further insights into the underlying mechanisms.

The organization of this paper is as follows: In Section II, we describe the model and elaborate on the formalism. In Section III, we present the numerical results. In Section IV, we discuss the origins of the numerical results. Finally, we provide a brief summary of our work in Section V.

\section{\label{sec2.mod}MODEL AND FORMALISM}

Our analysis is anchored in the Hamiltonian, which is composed of three components: the tight-binding term, the on-site interaction term, and the off-site pairing interaction term,
\begin{equation}
	H = H_t + H_{\text{int}} + H_p.
\end{equation}

The tight-binding term \( H_t \) includes both the hopping and chemical potential contributions,
\begin{equation}
	H_t = -\sum_{l,l'} \sum_{\mathbf{i}\mathbf{j}\tau\tau'\sigma} t^{l,l'}_{\mathbf{i}\mathbf{j}\tau\tau'} c^{l\dagger}_{\mathbf{i}\tau\sigma} c^{l'}_{\mathbf{j}\tau'\sigma} - \mu_0 \sum_{l} \sum_{\mathbf{i}\tau\sigma} c^{l\dagger}_{\mathbf{i}\tau\sigma} c^{l}_{\mathbf{i}\tau\sigma},
\end{equation}
where \( l \)/\( l' \), \( \tau \)/\( \tau' \), and \( \sigma \) represent layer, orbital, and spin indices, respectively. 

\( H_{\text{int}} \) is the on-site interaction term, articulated as,
\begin{equation}
	\begin{aligned}
		H_{\text{int}} &= U \sum_{l} \sum_{\mathbf{i}\tau} n_{\mathbf{i}\tau\uparrow}^{l} n_{\mathbf{i}\tau\downarrow}^{l} + U' \sum_{l} \sum_{\mathbf{i},\tau<\tau'} n_{\mathbf{i}\tau}^{l} n_{\mathbf{i}\tau'}^{l} \\
		&\quad + J \sum_{l} \sum_{\mathbf{i},\tau<\tau',\sigma\sigma'} c^{l\dagger}_{\mathbf{i}\tau\sigma} c^{l\dagger}_{\mathbf{i}\tau'\sigma'} c_{\mathbf{i}\tau\sigma'}^{l} c_{\mathbf{i}\tau'\sigma}^{l}\\&\quad + J' \sum_{l} \sum_{\mathbf{i},\tau\neq\tau'} c^{l\dagger}_{\mathbf{i}\tau\uparrow} c^{l\dagger}_{\mathbf{i}\tau\downarrow} c_{\mathbf{i}\tau'\downarrow}^{l} c_{\mathbf{i}\tau'\uparrow}^{l},
\end{aligned}
\end{equation}
with \( n_{i\alpha} = n_{\alpha\uparrow} + n_{\alpha\downarrow} \), and \( U \), \( U' \), \( J \), and \( J' \) representing the onsite orbital intra- and inter-orbital repulsion, the onsite Hund's coupling, and the pairing hopping term, respectively.  The relationships between these parameters are \( U = U' + 2J \) and \( J = J' \).

The off-site pairing potential \( H_p \) accounts for the superconducting pairing,
\begin{equation}
	H_p = \sum_{l,l'} \sum_{\mathbf{ij}\tau} V_{\mathbf{ij}\tau}^{l,l'} \left( c_{\mathbf{i}\tau\uparrow}^{l\dagger} c_{\mathbf{j}\tau\downarrow}^{l'\dagger} c_{\mathbf{i}\tau\uparrow}^{l} c_{\mathbf{j}\tau\downarrow}^{l'} + c_{\mathbf{i}\tau\downarrow}^{l\dagger} c_{\mathbf{j}\tau\uparrow}^{l'\dagger} c_{\mathbf{i}\tau\downarrow}^{l} c_{\mathbf{j}\tau\uparrow}^{l'} \right).
\end{equation}

At the meanfield level, we define the mean-field order parameters for intra- and interlayer pairings as \( \Delta_{\mathbf{ij}\tau}^{l,l} = \frac{V_{\mathbf{ij}\tau}^{l,l}}{2} \langle c_{\mathbf{i}\tau\uparrow}^{l} c_{\mathbf{j}\tau\downarrow}^{l} - c_{\mathbf{i}\tau\downarrow}^{l} c_{\mathbf{j}\tau\uparrow}^{l} \rangle \), \( \Delta_{\mathbf{ii}\tau}^{l,l'} = \frac{V_{\mathbf{ii}\tau}^{l,l'}}{2} \langle c_{\mathbf{i}\tau\uparrow}^{l} c_{\mathbf{i}\tau\downarrow}^{l'} - c_{\mathbf{i}\tau\downarrow}^{l} c_{\mathbf{i}\tau\uparrow}^{l'} \rangle \), respectively. The superconducting pairing term is then expressed as,
\begin{equation}
	H_p = -\sum_{l,l'} \sum_{\mathbf{ij}\tau} \left( \Delta^{l,l'}_{\mathbf{ij}\tau} c^{l\dagger}_{\mathbf{i}\tau\uparrow} c^{l'\dagger}_{\mathbf{j}\tau\downarrow} + \text{H.c.} \right).
\end{equation}

In momentum space, the basis vector is defined as \( \Psi(\mathbf{k}) = (u_{\mathbf{k}}, v_{\mathbf{k}}) \). $u_{\mathbf{k}}$ and $v_{\mathbf{k}}$ represent the electron and hole components, respectively, expressed as,
\begin{eqnarray}
	\begin{aligned}
		&u_{\mathbf{k}} = (c^{1}_{\mathbf{k}1\uparrow}, c^{1}_{\mathbf{k}2\uparrow}, c^{2}_{\mathbf{k}1\uparrow}, c^{2}_{\mathbf{k}2\uparrow})^T, \\
		&v_{\mathbf{k}} = (c^{1\dagger}_{-\mathbf{k}1\downarrow}, c^{1\dagger}_{-\mathbf{k}2\downarrow}, c^{2\dagger}_{-\mathbf{k}1\downarrow}, c^{2\dagger}_{-\mathbf{k}2\downarrow})^T.
\end{aligned}
\end{eqnarray}
The bare Hamiltonian can be expressed as \( H_0  = \sum_{\mathbf{k}} \Psi^{\dagger}_{\mathbf{k}} \hat{H}_{\mathbf{k}} \Psi_{\mathbf{k}} \), where \(\hat{H}_{\mathbf{k}} \) is an \( 8 \times 8 \) matrix, expressed as,
\begin{equation}
	\hat{H}_\textbf{k}=
	\begin{bmatrix}
		\hat{H}_0(\textbf{k})  & \hat{\Delta}(\textbf{k})  \\
		\hat{\Delta}^{\dagger}(\textbf{k}) & -\hat{H}_0(\textbf{k})
	\end{bmatrix}.
\end{equation}
$\hat{H}_0(\textbf{k})$ is a $4\times 4$ matrix which is obtained from the tight-binding term $H_t$.  

 $\hat{\Delta}(\textbf{k})$ is the $4\times 4$ superconducting pairing matrix, expressed as,
 \begin{equation}
	\hat{\Delta}(\textbf{k})=
	\begin{pmatrix}
		\Delta_{x\parallel} ({\bf k}) & 0 & \Delta_{x\perp} & 0  \\
		0 & \Delta_{z\parallel} ({\bf k}) & 0 & \Delta_{z\perp} \\
		\Delta_{x\perp} & 0 & \Delta_{x\parallel} ({\bf k}) & 0 \\
		 0& \Delta_{z\perp} & 0 &  \Delta_{z\parallel} ({\bf k})
	\end{pmatrix}.
\end{equation}
The diagonal elements in the above pairing matrix represent the intralayer pairing and the off-diagonal ones represent the interlayer pairing. 
 For intralayer pairing, we concentrate on the nearest-neighbor interaction, represented by $ V_{\tau\parallel} = V^{l,l}_{\mathbf{i}\mathbf{j}\tau} $. Then the self-consistent calculation support the extended $s$-wave pairing symmetry with $\Delta_{\tau\parallel} ({\bf k})=2\Delta_{\tau\parallel} (\cos k_x+\cos k_y)$.
 The pairing order parameter is given by~\cite{PhysRevB.108.174501},
\begin{equation}
	\Delta_{\tau\parallel} = \frac{V_{\tau\parallel}}{4N} \sum_{n\mathbf{k}} \left( \cos \mathbf{k}_x + \cos \mathbf{k}_y \right) u_{\tau n\mathbf{k}}^{l*} v_{\tau n\mathbf{k}}^{l} \tanh \frac{\beta E_{n\mathbf{k}}}{2}.
\end{equation}

For interlayer pairing, the interaction is characterized by \( V_{\tau\perp} = V^{l,l'}_{\mathbf{i}\mathbf{i}\tau} \), the pairing order parameter can be similarly determined as,
\begin{equation}
	\Delta_{\tau\perp} = \frac{V_{\tau\perp}}{2N} \sum_{n\mathbf{k}} u_{\tau n\mathbf{k}}^{l*} v_{\tau n\mathbf{k}}^{l'} \tanh \frac{\beta E_{n\mathbf{k}}}{2}.
\end{equation}

In the absence of interaction terms, the bare spin susceptibility of the superconducting state, encompassing both normal and anomalous contributions, is formulated as a $16\times 16$ matrix, with the elements being expressed as~\cite{PhysRevLett.91.037002,PhysRevB.69.224514,PhysRevB.103.174513},
\begin{equation}
	\begin{aligned}
		\chi^{l_1l_2l_3l_4}_0(\textbf{q},\omega) &= \frac{1}{N} \sum_{\textbf{k}ij} \left[ u_{l_1 i}(\textbf{k}) u_{l_2 i}(\textbf{k}) u_{l_3 j}(\textbf{k}+\textbf{q}) u_{l_4 j}(\textbf{k}+\textbf{q}) \right. \\
		&\quad \left. + u_{l_1 i}(\textbf{k}) u_{l_2+4, i}(\textbf{k}) u_{l_3 j}(\textbf{k}+\textbf{q}) u_{l_4+4, j}(\textbf{k}+\textbf{q}) \right] \\
		&\quad \times \frac{f(E_j(\textbf{k}+\textbf{q})) - f(E_i(\textbf{k}))}{\omega - E_j(\textbf{k}+\textbf{q}) + E_i(\textbf{k}) + i\delta},
\end{aligned}
\end{equation}
where \( u_{ij}(\textbf{k}) \) and \( E_i(\textbf{k}) \) are the eigenvectors and eigenvalues of the Hamiltonian matrix \( \hat{H}_{\bf k} \), obtained by diagonalizing the Hamiltonian matrix. $f(E)$ is the Fermi-Dirac function, and $l_i$ are the orbital and layer indices ranging from $1$ to $4$. $u_{l_i i}$ and $u_{l_{i+4},i}$ are derived from the electron and hole operators, respectively. 

The renormalized spin susceptibility \( \chi(\textbf{q},\omega) \)  is derived using the random phase approximation (RPA),
\begin{equation}
	\chi^{RPA}_{l_1l_2l_3l_4}(\textbf{q},\omega) = \left[ \hat{\chi}_0(\textbf{q},\omega) \left( \hat{I} -\alpha \hat{U} \hat{\chi}_0(\textbf{q},\omega) \right)^{-1} \right]_{l_1l_2l_3l_4},
\end{equation}
where \( \hat{I} \) is a \( 16 \times 16 \) unit matrix. Considering on-site interactions, the elements of the \( \hat{U} \) matrix can be expressed as~\cite{Kemper_2010,PhysRevB.96.014515},
\begin{equation}
	\hat{U}_{l_1l_2l_3l_4} = 
	\begin{cases}
		{U}, & \text{if } l_1 = l_2 = l_3 = l_4, \\
		{U'}, & \text{if } l_1 = l_3 \neq l_2 = l_4, \\
		{J}, & \text{if } l_1 = l_2 \neq l_3 = l_4, \\
		{J'}, & \text{if } l_1 = l_4 \neq l_2 = l_3.
	\end{cases}
\end{equation}
We here include an additional factor $\alpha$ in the RPA approximation, as is seen in Eq.(10). The existence of this additional factor $\alpha$ within the RPA framework is generally unitized. The effective interaction within the RPA framework is typically much smaller than the original on-site repulsion, ensuring the validity of the RPA approach~\cite{HiroyukiYamase1999,PhysRevLett.82.2915}. For the one-band $t-J$ model, which describes cuprate superconductors, it is suggested that the renormalized interaction in the RPA should be scaled by a factor $\alpha = 0.34$ to align with antiferromagnetic instability~\cite{PhysRevLett.82.2915}. The value of $\alpha$ can be estimated through the condition for magnetic instability, namely, the magnetic instability occurs when the pole condition of the real part of the zero-energy RPA factor, $\det\lvert \hat{I} - \alpha\hat{U}\text{Re}\hat{\chi}_0(q,0) \rvert = 0$, is satisfied.

The physical spin susceptibility is obtained by summing the diagonal elements of the \( \hat{\chi} \) corresponding to \( l_4 = l_1 \) and \( l_3 = l_2 \)~\cite{Kemper_2010,PhysRevB.96.014515}, expressed as,
\begin{equation}
	\chi^{RPA}(\textbf{q},\omega) = \sum_{l_1,l_2} \chi^{RPA}_{l_1l_2l_2l_1}(\textbf{q},\omega).
\end{equation}

In the following presented results, the hopping constants in Eq.(2) are sourced from Ref.~\cite{PhysRevLett.131.126001} with the energy unit being eV. The interaction parameter values are adopted from~\cite{PhysRevLett.131.206501}, with \( U = 3.7  \), \( U' = 2.5  \), and \( J = J' = 0.6  \). Unless otherwise specified, the default parameters include \( T = 0.001 \), \( \delta = 0.002 \), and the summation over the wave vector \(\mathbf{k}\) is performed by dividing the Brillouin zone into a \(160 \times 160\) lattice grid.

\section{\label{sec3.res}RESULTS}

\begin{figure} 
	\centering 
	\includegraphics[width = 8cm]{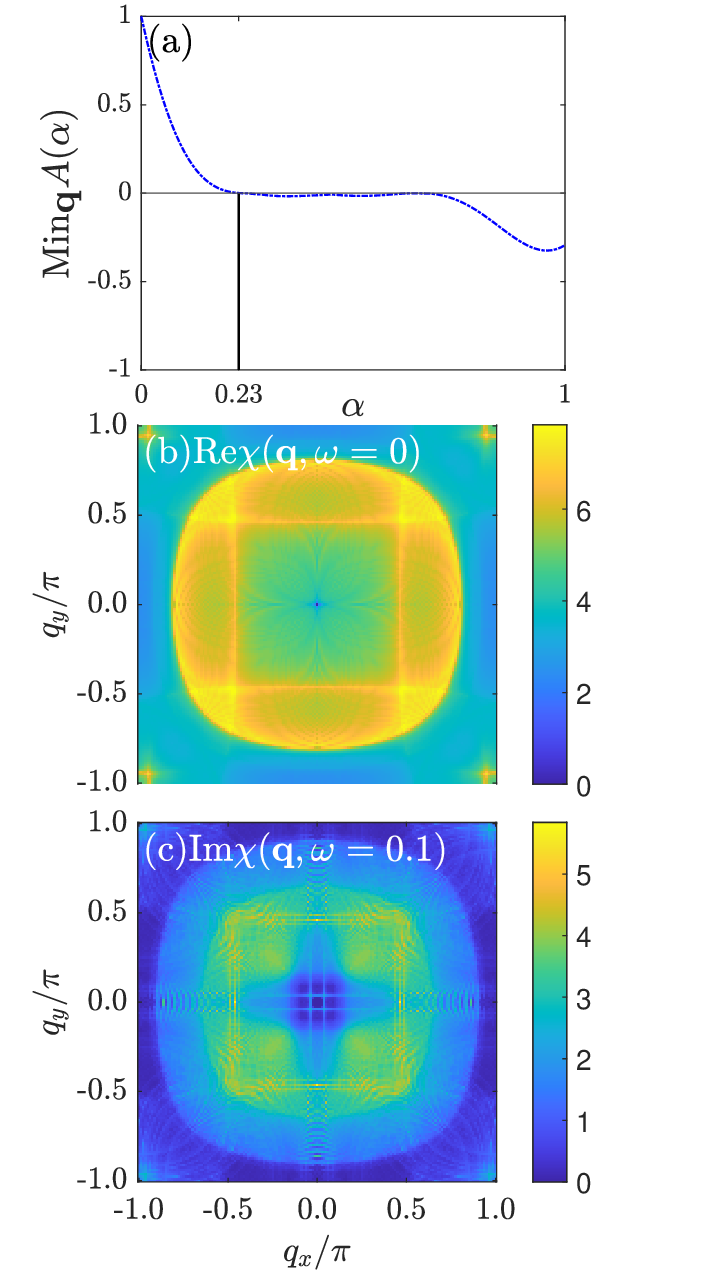} 
\caption{Numerical results of the normal state spin excitations.(a) Depiction of the minimum RPA factor value as a function of the additional factor $\alpha$ with $\omega=0$. (b) Illustration of the real part of the zero-energy spin susceptibility plotted against momentum ${\bf q}$. (c) Display of the imaginary part of the spin susceptibility as a function of momentum ${\bf q}$ with $\omega=0.1$.
}
	\label{fig.1} 
\end{figure}

Our studies commence with an analysis of spin excitations in the normal state, achieved by setting the superconducting gap $\Delta$ to zero. We initially validate the effective interaction within the RPA framework, where the effective interaction is typically much smaller than the original on-site repulsion, ensuring the validity of the RPA approach~\cite{HiroyukiYamase1999,PhysRevLett.82.2915}. For the one-band $t-J$ model, which describes cuprate superconductors, it is suggested that the renormalized interaction in the RPA should be scaled by a factor $\alpha = 0.34$ to align with antiferromagnetic instability~\cite{PhysRevLett.82.2915}. We also incorporated an additional factor into the RPA factor of the spin susceptibility [Eq.(10)]. The condition for magnetic instability is met when the pole condition of the real part of the zero-energy RPA factor, $\det\lvert \hat{I} - \alpha\hat{U}\text{Re}\hat{\chi}_0(q,0) \rvert = 0$, is satisfied.

Fig. \ref{fig.1}(a) represents the minimum value of the RPA factor at zero energy as a function of the additional factor $\alpha$. It is evident that for $\alpha$ exceeding the critical threshold of 0.23, the minimum value of the RPA factor, $A(q,0)$, becomes negative, indicating magnetic instability. To avoid this instability in subsequent calculations, we use $\alpha = 0.2$, and we verified through numerical analysis that the results remain stable even with different values of $\alpha$~\cite{supp}.

The real part of the renormalized spin susceptibility as a function of momentum ${\bf q}$ is plotted in Fig. \ref{fig.1}(b), revealing that maximum spin excitations peak at an incommensurate wave vector near ${\bf q} = (0.5\pi, 0.5\pi)$, with the highest value observed at an incommensurate momentum of $\mathbf{q} = (0.44\pi, 0.44\pi)$. The imaginary part of the spin susceptibility at the energy $\omega = 0.1$ is depicted in Fig. \ref{fig.1}(c), indicating that dominant spin excitations occur near the wave vector ${\bf q} = (0.5\pi, 0.5\pi)$, with the maximum value at an incommensurate momentum ${\bf q} = (0.46\pi, 0.46\pi)$. It is important to note that the degree of incommensurability is weakly dependent on the factor $\alpha$, as detailed in the supplementary material~\cite{supp}.

Our numerical outcomes corroborate prior experimental and theoretical discoveries. Experimental evidence of magnetic order near the momentum $(0.5\pi, 0.5\pi)$ was observed in the parent compound under ambient pressure \cite{chen2024electronic}. Numerically, the $(0.5\pi, 0.5\pi)$ in-plane modulation was also identified using the RPA approach~\cite{PhysRevLett.131.126001}.

Before delving into the spin excitations within the superconducting state, we first investigate the superconducting pairing state. In our current study, we rely on the hopping parameters derived from the first-principles calculations as reported in Ref.~\cite{PhysRevLett.131.126001}. Specifically, the interlayer hopping constant for the $d_{z^2}$ orbital ($t^z_{\perp}$) and the nearest-neighbor intralayer hopping constant for the $d_{x^2-y^2}$ orbital ($t^x_{\parallel}$) are identified as the most significant, with values of $t^x_{\parallel} = 0.483$ and $t^z_{\perp} = 0.635$, respectively. Other hopping constants are considerably smaller in comparison. It is widely accepted that pairing potentials are strongly influenced by these hopping constants. Given the assumption that pairing interactions originate from superexchange interactions, the ratio of the pairing potentials is estimated to be $V_{x\parallel}/V_{z\perp} \approx (t^x_{\parallel}/t^z_{\perp})^2 \approx 0.58$. This ratio is consistent with those proposed in previous theoretical studies~\cite{PhysRevLett.132.146002,wu2024}.

\begin{figure} 
	\centering 
	\includegraphics[width = 8cm]{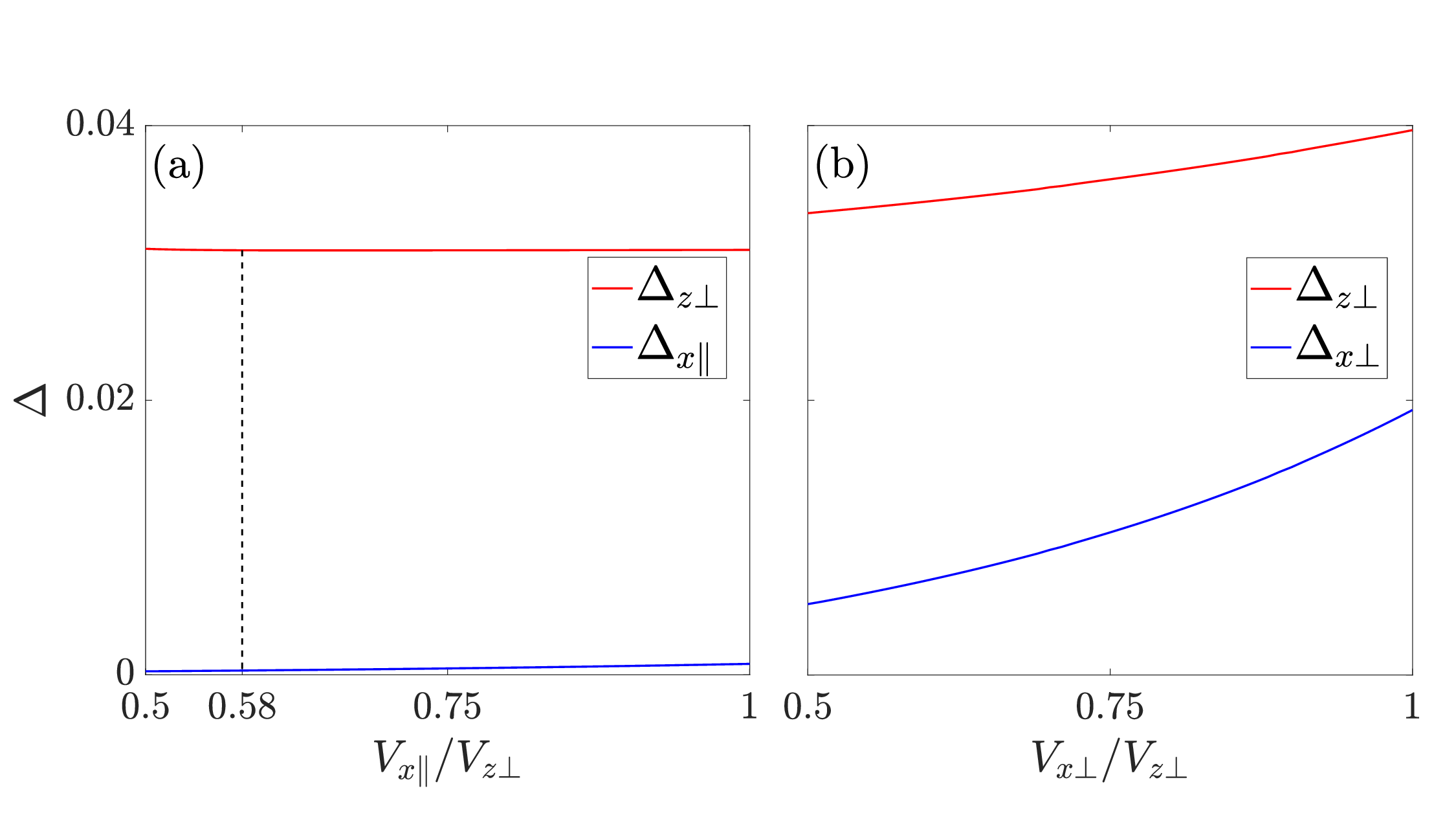} 
	\caption{Variation of order parameters with pairing potential
		from self-consistent calculations. }
	\label{fig.2} 
\end{figure}

By setting the interlayer pairing strength for the $d_{z^2}$ orbital to $V_{z\perp} = 0.6$, we applied a self-consistent approach to investigate the behavior of order parameters in relation to the intralayer pairing strength, $V_{x\parallel}$. As depicted in Fig. \ref{fig.2}(a), with a ratio of $V_{x\parallel}/V_{z\perp} = 0.58$, the intralayer pairing magnitude is found to be nearly negligible. Further increases in $V_{x\parallel}$ lead to a slight enhancement in the intralayer order parameters; nonetheless, they remain substantially lower than the interlayer parameter, even when $V_{x\parallel}$ matches $V_{z\perp}$. Our numerical findings robustly suggest that interlayer pairing plays a predominant role in La$_3$Ni$_2$O$_7$, highlighting its importance over intralayer pairing in shaping the superconducting properties of this material.

Experimentally, while there is no direct observation to confirm the dominant interlayer pairing, evidence supporting a dominant interlayer superexchange interaction has been inferred from resonant inelastic X-ray scattering and neutron scattering measurements~\cite{xie2024neutron,chen2024electronic}.

\begin{figure} 
	\centering 
	\includegraphics[width = 8cm]{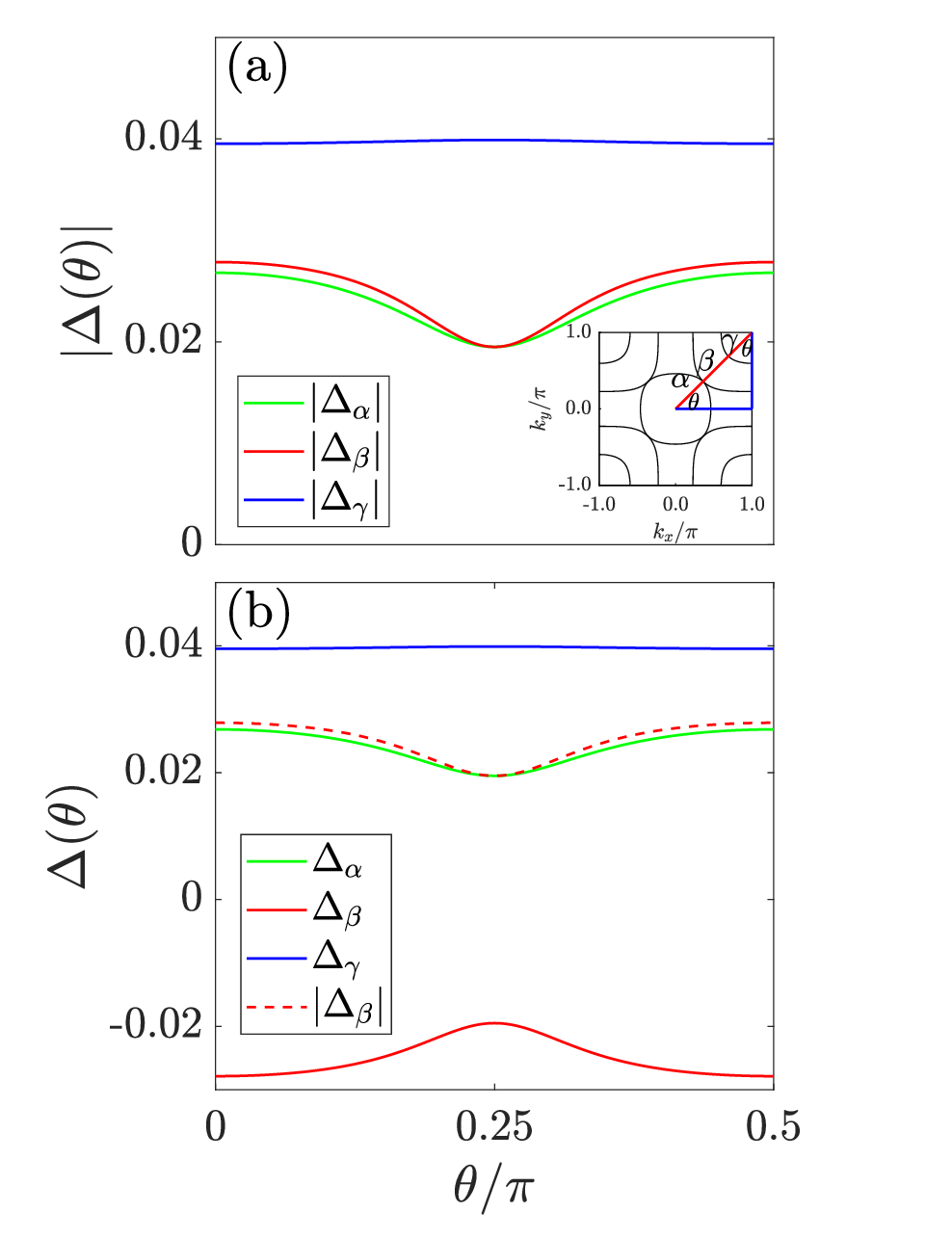} 
	\caption{(a) Magnitudes of the energy gap along different Fermi pockets, derived from quasiparticle energy minima.
		(b) Energy gaps with sign information along the Fermi pockets, calculated from the band-basis pairing matrix.
	}
	\label{fig.3} 
\end{figure}

Despite the small interlayer hopping constant for the $d_{x^2-y^2}$ orbital, which effectively eliminates direct superexchange interactions~\cite{PhysRevLett.131.126001}, significant Hund's coupling enables the $d_{z^2}$ orbital to impart an interlayer pairing potential onto the $d_{x^2-y^2}$ orbital. This mechanism facilitates interlayer pairing even when the direct superexchange is minimal~\cite{arXiv2310.02915,PhysRevB.108.174511,arXiv2311.05491,PhysRevLett.132.146002,arXiv2402.07449}. Our numerical analysis, accounting for an effective interlayer pairing potential $V_{x\perp}$ for the $d_{x^2-y^2}$ orbital, outlines the behavior of order parameters as $V_{x\perp}$ varies from 0.3 to 0.6, as depicted in Fig.~\ref{fig.2}(b). As $V_{x\perp}$ increases, the pairing order parameter for the $d_{x^2-y^2}$ orbital is not only initiated but also strengthened. Simultaneously, there is a slight increase in the pairing order parameter of the $d_{z^2}$ orbital. This result suggests that Hund's coupling may have the potential to enhance the superconducting transition temperature. In subsequent analyses, we set the intralayer pairing interaction to zero and consider interlayer pairing involving both orbitals, adopting $V_{x\perp} = V_{z\perp} = 0.6$ for illustrative purposes.

The superconducting gaps across various Fermi pockets in momentum space are illustrated in Fig. \ref{fig.3}. Fig.~\ref{fig.3}(a) presents the superconducting gap derived from the minimum quasiparticle energy along the normal-state Fermi surface in the superconducting state, focusing solely on the gap magnitude without phase information. The superconducting gap is anisotropic and associated with the orbital contributions to the Fermi surface~\cite{PhysRevLett.131.126001,PhysRevB.110.L060506}. Specifically, as indicated in Ref.~\cite{PhysRevLett.131.126001}, the $\gamma$ Fermi pocket is primarily contributed by the $d_{z^2}$
orbital. The energy gap around the $\gamma$ Fermi pocket is nearly uniform, equaling $\Delta_{z\perp}$. The $\alpha$ and $\beta$ pockets are contributed by both $d_{x^2-y^2}$
and $d_{z^2}$ orbitals, with orbital weights varying with $\theta$, resulting in anisotropic energy gaps around these pockets. Along the diagonal direction with $\theta=0.25$, the $d_{x^2-y^2}$
orbital dominates, contributing to the Fermi surface. Consequently, for these pockets, the energy gap equals $\Delta_{x\perp}$
​when the Fermi momentum is along the diagonal direction. As the Fermi momentum deviates from the diagonal direction, the $d_{z^2}$
orbital weight increases, leading to larger gap magnitudes. The entire Fermi surface is fully gapped, with a minimum gap $\Delta_m=0.02$, primarily determined by the pairing potential in the $d_{x^2-y^2}$
​orbital. Our numerical results for the gap magnitudes around the $\alpha$ and $\beta$ pockets are qualitatively consistent with experimental measurements of the superconducting gaps around these pockets~\cite{shen2025}.


To determine the phase information of the superconducting gap around each Fermi surface, the pairing matrix must be transformed from the orbital basis to the band basis. This is achieved by setting the intralayer pairing functions to zero, i.e., $\Delta_{\tau\parallel}({\bf k})=0$ in Eq.(8),
 to obtain the pairing matrix with merely the interlayer pairing in the orbital basis. Subsequently, the pairing matrix in the band basis, $\hat{\Delta}_B({\bf k})$, is calculated using $\hat{\Delta}_B({\bf k})=\hat{V}^\dagger({\bf k}) \hat{\Delta}({\bf k}) \hat{V}({\bf k})$, where  $\hat{V}({\bf k})$  is the eigenvector matrix of the
$4\times 4$ normal state matrix $\hat{H}_0({\bf k})$. The pairing function around the normal-state Fermi surface is then extracted from the diagonal elements of $\hat{\Delta}_B({\bf k})$. Fig.~\ref{fig.3}(b) replots the numerical results of the pairing function along the normal-state Fermi surface based on the pairing matrix in the band basis. As illustrated, the pairing gaps along the $\gamma$ and $\alpha$ Fermi pockets are positive, while those along the $\beta$ pocket are negative. Our findings reveal that interlayer $s$-wave pairing leads to an effective $s_{\pm}$ symmetry.
Notably, the gap magnitudes in Fig.~\ref{fig.3}(b) are consistent with those in Fig.~\ref{fig.3}(a).

\begin{figure} 
	\centering 
	\includegraphics[width = 8cm]{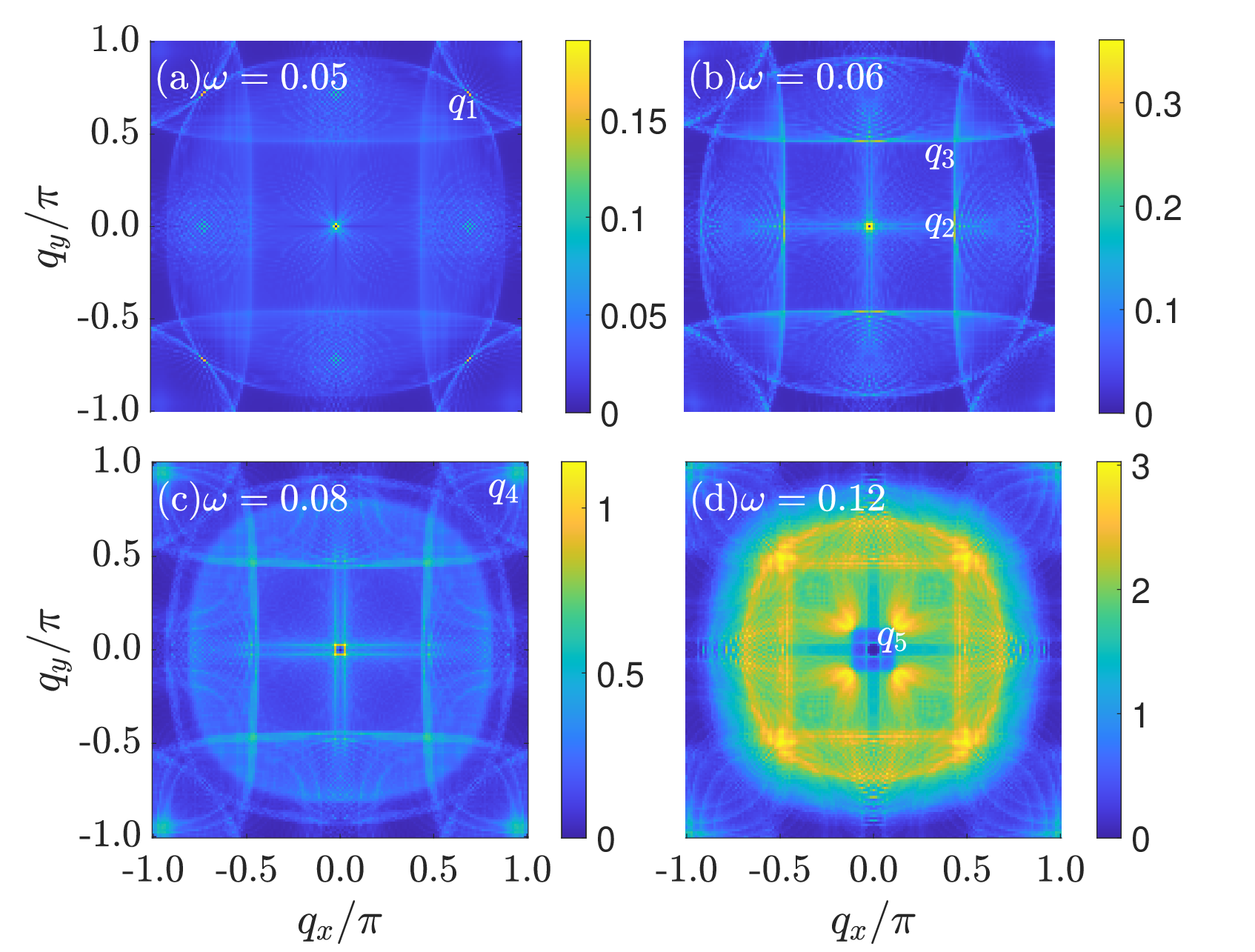} 
	\caption{Imaginary parts of the spin susceptibility in the superconducting state as a function of the momentum ${\bf q}$ with different energies. }
	\label{fig.4} 
\end{figure}

We now present the numerical results of spin excitations in the superconducting state. The intensity plots depicting the imaginary part of spin susceptibility as a function of momentum for various energies are shown in Fig.~\ref{fig.4}. These plots reveal that spin excitations emerge around the momentum $(0,0)$ and $\mathbf{q}_1-\mathbf{q}_5$. The spin excitations exhibit a strong dependence on energy. At a low energy of $\omega=0.05$, only the spin excitations around $(0,0)$ and $\mathbf{q}_1$ are present. As the energy increases to $0.06$, spin excitations emerge around the momenta $\mathbf{q}_2 \approx (\pi/2,0)$ and $\mathbf{q}_3 \approx (\pi/2,\pi/2)$. Upon further energy increase to $0.08$ and $0.12$, additional peaks of spin excitations at momenta $\mathbf{q}_4$ and $\mathbf{q}_5$ appear.

\begin{figure} 
	\centering 
	\includegraphics[width = 8cm]{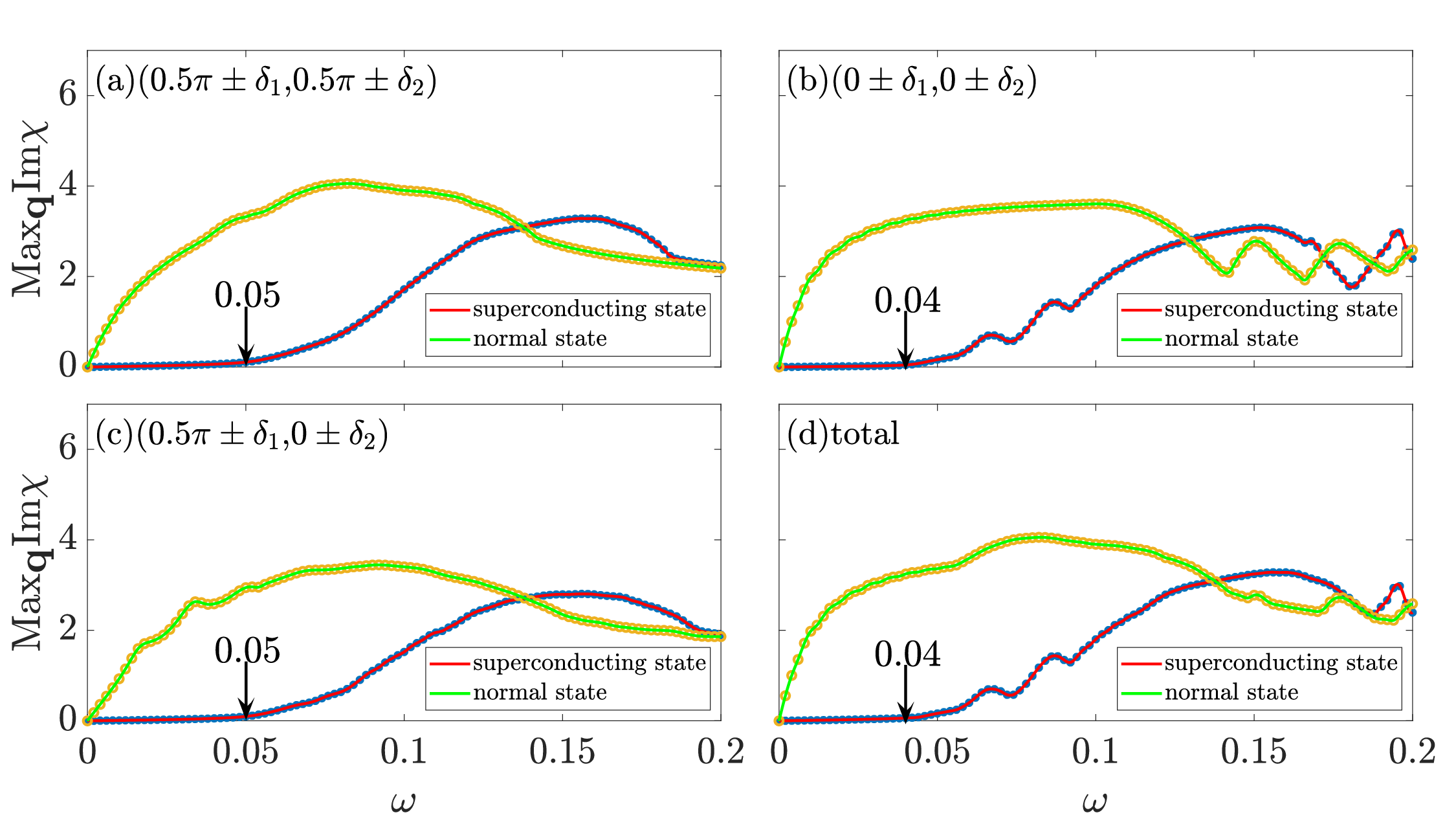} 
	\caption{The maximum value of the imaginary part of the susceptibility, Im$\chi({\bf q}, \omega)$, as a function of energy across various momenta, with a quasiparticle damping of $\delta=0.005$. The Brillouin zone is discretized into a $640 \times 640$ lattice for the summation over ${\bf k}$. }
	\label{fig.5} 
\end{figure}

Spin excitations in the superconducting state encompass several contributing parts. The spin susceptibility is peaked at different momenta, primarily around $(0,0)$, $(\pi/2,0)$, and $(\pi/2,\pi/2)$, as depicted in Fig.~\ref{fig.4}. 
The numerical results of the maximum imaginary parts of spin susceptibility around these three momenta, as functions of energy, are presented in Figs.~\ref{fig.5}(a)-\ref{fig.5}(c). The maximum spin susceptibility throughout the entire Brillouin zone is displayed in Fig.~\ref{fig.5}(d). It is noticeable that in the superconducting state, a spin gap of about $0.04$ exists (about $2\Delta_m$).

The spin excitations around $(\pi/2,\pi/2)$ and $(\pi/2,0)$ come into play when the energy exceeds $0.05$. At lower energies, the spin excitation in the superconducting state is significantly less than in the normal state, due to the presence of the superconducting gap. Our findings suggest that there is no spin resonance mode in the superconducting state. As energy increases beyond $0.18$, the spin excitation in the superconducting state aligns closely with that in the normal state.

\section{\label{sec4.res}Discussion}

We first focus on the incommensurate patterns of spin excitations at various momenta. These patterns can be understood by examining the bare spin susceptibility, which is primarily influenced by particle-hole excitations and is given by the expression $\text{Im}\chi_0({\bf q},\omega) \propto \sum_{\bf k}\delta(\omega-E({\bf k})-E({\bf k}+{\bf q}))$.

\begin{figure} 
	\centering 
	\includegraphics[width = 8cm]{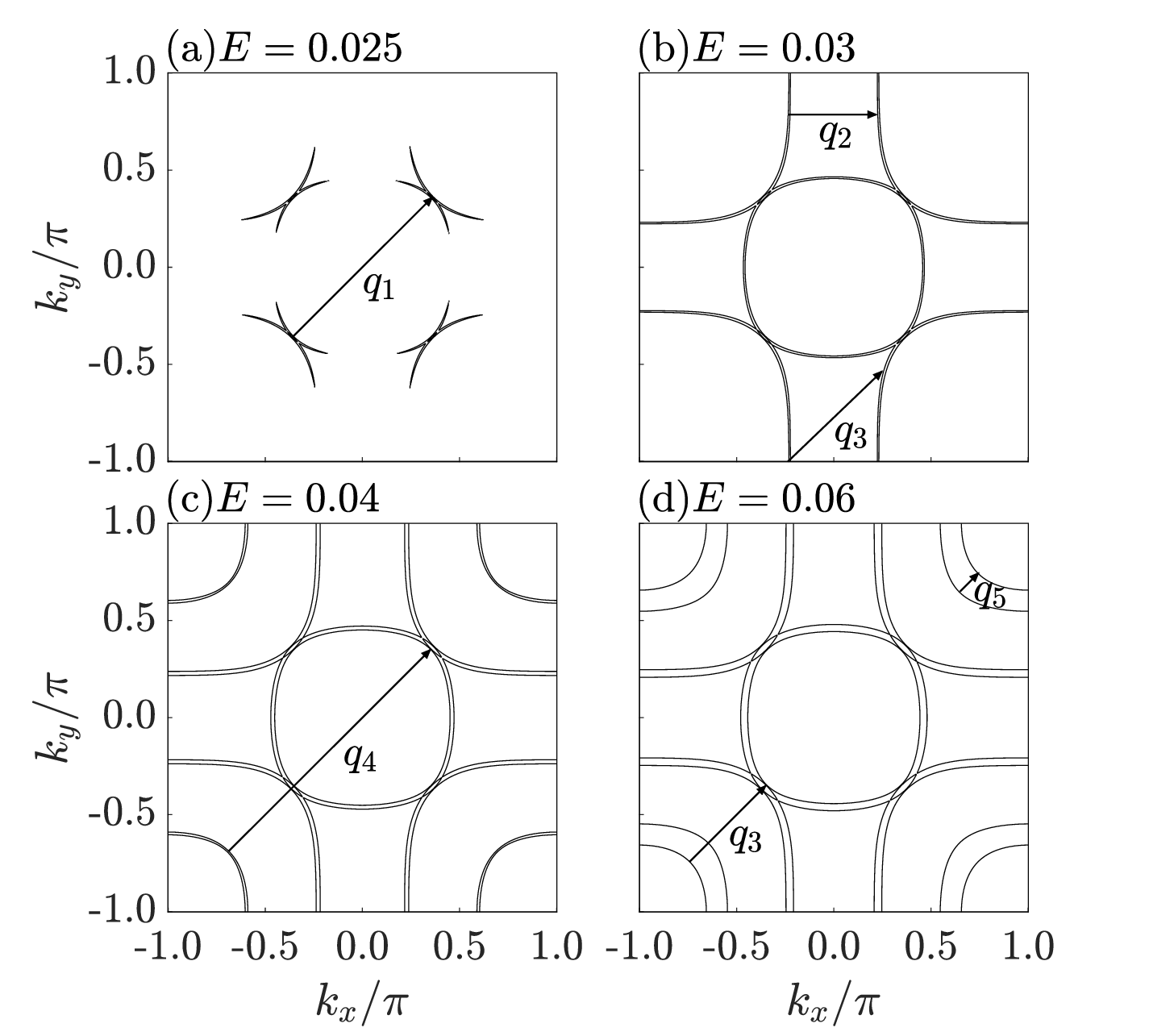} 
\caption{The constant energy contours with different energies. The arrows indicate the nesting vectors.}
	\label{fig.6} 
\end{figure}

To clarify the spin excitation at a specific energy $\omega_0$, it is essential to investigate the nesting wave vector between the energy contours at $E = \omega_0/2$~\cite{PhysRevLett.82.2915}. The contour plots of energy contours at different energies are shown in Fig.~\ref{fig.6}, with the nesting wave vectors indicated. These are found to be in good agreement with the numerical results of spin excitations presented in Fig.~\ref{fig.4}. It is important to note that the nesting wave vectors ${\bf q_1}$ to ${\bf q_5}$ are primarily determined by the normal state Fermi surface.
 In the normal state, the dominant spin excitations also appear near these momenta, as shown in Fig.~\ref{fig.1}. The nesting vector ${\bf q_3}$ has contributions from two parts, as seen in Figs.~\ref{fig.6}(b) and \ref{fig.6}(d). Consequently, the normal state spin excitations near the momentum ${\bf q_3}$ [around $(\pi/2,\pi/2)$] are relatively strong, as shown in Fig.~\ref{fig.1}. 

In the superconducting state, at low energies, the dominance of a particular nesting vector is influenced by the low-energy quasiparticle excitations, which are closely related to the superconducting pairing symmetry.
For the interlayer pairing term considered here, as shown in Fig.~\ref{fig.3}, the minimum energy gap occurs at the diagonal direction with $\theta=0.25\pi$. Consequently, the low-energy contour with $\omega=0.02$ first emerges around the diagonal direction, as depicted in Fig.~\ref{fig.6}(a). In this case, only the nesting vector ${\bf q_1}$ is present. As the energy increases to $\omega=0.03$, which exceeds the maximum energy gap along the $\alpha$ and $\beta$ Fermi pockets, the energy contours surround the entire $\alpha$ and $\beta$ Fermi pockets, and the nesting vectors ${\bf q_2}$ and ${\bf q_3}$ appear [Fig.~\ref{fig.6}(b)]. When the energy further increases to values larger than the energy gap around the $\gamma$ pocket, additional energy contours surrounding the $\gamma$ pocket emerge, leading to the appearance of nesting vectors ${\bf q_4}$ and ${\bf q_5}$ [Fig.~\ref{fig.6}(c) and \ref{fig.6}(d)].

In fact, the low-energy incommensurate spin excitations are determined by the low-energy energy contours, which strongly depend on the pairing functions. Consequently, the low-energy incommensurate spin excitations also play a significant role in determining the pairing symmetry \cite{supp}.

The absence of a spin resonance mode in the superconducting state of 
La$_3$Ni$_2$O$_7$ can be understood through the interplay between Cooper pairing symmetry and spin susceptibility. In unconventional superconductors, the bare spin excitations without the RPA correction are modulated by the coherence factor, which depends on the relative signs of the superconducting gaps $\Delta_{\mathbf{k}}$ and $\Delta(\mathbf{k}+\mathbf{Q})$ at momenta connected by the nesting vector ${\bf Q}$~\cite{PhysRevLett.82.2915,PhysRevB.58.2895,PhysRevB.62.640}. 
A sign reversal with
$\Delta_{\mathbf{k}}=-\Delta(\mathbf{k}+\mathbf{Q})$ 
  enhances low-energy spin excitations, generating a step-like increase in the imaginary part of the bare spin susceptibility at the spin gap edge. By the Kramers-Kronig relation, this produces a logarithmic divergence in the real part 
of the bare spin susceptibility, creating a pole in the RPA factor and enabling collective spin resonance modes. Conversely, when $\Delta_{\mathbf{k}}$ and $\Delta(\mathbf{k}+\mathbf{Q})$ share the same sign, spin excitations are suppressed at low energies.

For cuprate-based and iron-based high-T$_c$​
superconductors, spin-fluctuation-mediated pairing naturally favors a sign reversal of the gap between ${\bf k}$ and ${\bf k}+{\bf Q}$~\cite{Moriya_AdvPhys_2000_v49_p555,PhysRevLett.101.057003,Yao_2009}, making spin resonance a characteristic feature of these materials. La$_3$Ni$_2$O$_7$, however, has a more complex Fermi surface with multiple nesting vectors. While pairing symmetry has been extensively studied within a spin-fluctuation framework, with proposed d-wave and various $s_\pm$ pairing symmetries~\cite{PhysRevB.108.L140505,PhysRevLett.132.106002,arXiv2306.07275,PhysRevLett.131.236002,Zhang2024,PhysRevB.108.165141,PhysRevB.108.L201121}, a clear analytical form for $s_\pm$ pairing remains elusive.

Our analysis suggests that $s$-wave interlayer pairing dominates in La$_3$Ni$_2$O$_7$, yielding an effective 
$s_\pm$ symmetry in the band basis. However, a sign change in the order parameter does not automatically produce a spin resonance mode.~\cite{PhysRevLett.91.037002}.
Specifically, here
the sign-reversal condition $\Delta_{\mathbf{k}}=-\Delta(\mathbf{k}+\mathbf{Q})$ at the spin-gap edge 
required for spin resonance is not satisfied. As illustrated in Fig.~\ref{fig.6}(a),
low-energy quasiparticles emerge along diagonal momenta ${\bf k}$ and ${\bf k}+{\bf q_1}$, where $\Delta_{\mathbf{k}}=\pm\Delta(\mathbf{k}+\mathbf{q_1})$. Crucially, the sign alternation depends on whether ${\bf k}$ and $\mathbf{k}+\mathbf{q_1}$
  reside on the same Fermi pocket. The coexistence of positive and negative sign components prevents the characteristic step-like enhancement of bare spin susceptibility. 
   For nesting vectors ${\bf q_2}$ and ${\bf q_3}$ within the $\beta$ pocket [Fig.~\ref{fig.6}(b)], the gap retains the same sign, also precluding a step-like rise. Higher-energy vectors [the nesting vectors displayed in Fig.~\ref{fig.6}(c-d)] exceed the spin gap energy, rendering them irrelevant to low-energy resonance.  Thus, interlayer pairing fails to generate the requisite sign-reversal topology for spin resonance.

In contrast, intralayer unconventional pairings (e.g., extended 
$s$-wave or $d$-wave) inherently produce nodal lines via Fourier transformation of the pairing term. For nesting vectors ${\bf Q}$
crossing a nodal line, $\Delta_{\mathbf{k}}$ and $\Delta(\mathbf{k}+\mathbf{Q})$
naturally acquire opposite signs, fulfilling the resonance condition and enabling collective spin excitations~\cite{supp}. 
​

While this work elucidates key aspects of spin susceptibility in La$_3$Ni$_2$O$_7$, two critical challenges warrant further investigation. First, the current RPA framework for calculating spin susceptibility neglects self-energy effects in single-particle Green's functions—a simplification that may underestimate electronic correlation effects. Improving this calculation method by incorporating the effects of self-energy could provide a more accurate and comprehensive understanding of the spin susceptibility. Second, the pairing symmetry of La$_3$Ni$_2$O$_7$ remains unresolved, with competing proposals ranging from conventional $s$-wave to unconventional $d$-wave or mixed-symmetry states. Systematic exploration of alternative pairing scenarios—including orbital-selective pairing and complex gap structures with accidental nodes—should be pursued. Resolving these issues will not only advance our understanding of La$_3$Ni$_2$O$_7$ but also refine general methodologies for probing spin excitations in unconventional multi-orbital layered superconductors.

\section{\label{sec5.sum}SUMMARY}

In summary, our theoretical investigation of spin excitations in the bilayer nickelate superconductor La$_3$Ni$_2$O$_7$, based on a self-consistent mean-field approach, emphasizes the crucial role of interlayer pairing mechanisms in promoting superconductivity. The intriguing absence of a spin resonance mode, along with the emergence of energy-dependent incommensurate spin excitations, provides valuable insights into the fundamental superconducting mechanisms.

Our findings suggest that interlayer pairing dominates in La$_3$Ni$_2$O$_7$, leading to an effective $s_\pm$ symmetry. However, the sign change in the order parameter does not automatically result in a spin resonance mode.
The absence of a spin resonance mode can be attributed to the interplay between Cooper pairing symmetry and spin susceptibility. 
 This fact, combined with the observed incommensurate spin excitation structure—explained by the nesting effect seen in energy contours—suggests that spin excitations could serve as an essential diagnostic tool for examining the superconducting pairing mechanism. Collectively, these findings indicate a nuanced understanding of superconductivity in La$_3$Ni$_2$O$_7$, where interlayer pairing and energy contour nesting play decisive roles.

Our results, which differ from some prior studies, underscore the complexity of superconducting pairing in this material and call for further exploration. The energy-dependent incommensurate spin excitations, in particular, represent a fertile area for future research, potentially shedding more light on the relationship between spin dynamics and superconductivity in bilayer nickelates.

\begin{acknowledgments}
	This work was supported by the NSFC (Grant No.12074130).
\end{acknowledgments}


%

\renewcommand{\thesection}{S-\arabic{section}}
\setcounter{section}{0}  
\renewcommand{\theequation}{S\arabic{equation}}
\setcounter{equation}{0}  
\renewcommand{\thefigure}{S\arabic{figure}}
\setcounter{figure}{0}  
\renewcommand{\thetable}{S\Roman{table}}
\setcounter{table}{0}  
\onecolumngrid \flushbottom 
\newpage
\begin{center}\large \textbf{Supplemental material for spin excitations in bilayer La$_3$Ni$_2$O$_7$ superconductors with the interlayer pairing} \end{center}

\section{Spin excitation with different values of $\alpha$}

In the main text, we introduced an artificial factor $\alpha$ into the random phase approximation (RPA), with $\alpha = 0.2$ being the adopted value. We are now conducting a numerical study to determine if the qualitative nature of the results remains consistent when the value of $\alpha$ is varied. By considering various values of $\alpha$, we present the zero real part of the spin susceptibility in the normal state along the highly symmetric lines within the first Brillouin zone in Figure S1(a). Additionally, the imaginary part of the spin susceptibility in the superconducting state, with $\omega = 0.1$, is depicted in Figure S1(b). It is observed that as $\alpha$ decreases, the intensity of the spin susceptibility diminishes, yet the principal features of the spin excitation remain qualitatively unchanged. Specifically, the spin susceptibilities consistently exhibit peaks at certain momenta. The peak positions, which are indicative of the incommensurability, show minimal variation as $\alpha$ decreases.

\renewcommand \thefigure {S\arabic{figure}}

\begin{figure}[htb]
	\centering 
	\includegraphics[width = 14cm]{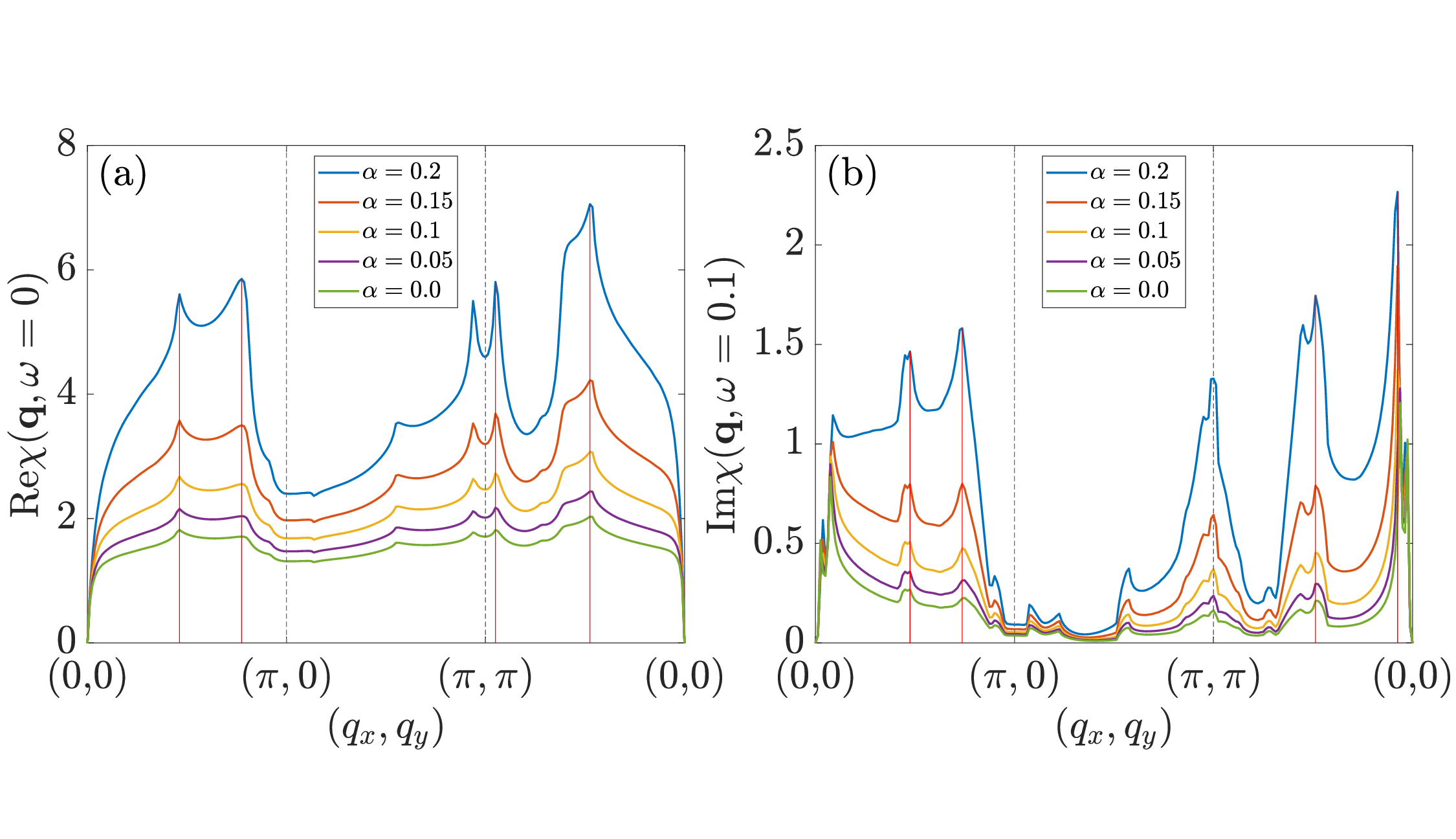} 
\caption{(a) Real part of the spin susceptibility in the normal state at zero energy, plotted as a function of momentum for various values of $\alpha$.
(b) Imaginary part of the spin susceptibility in the superconducting state at energy $\omega = 0.1$, plotted as a function of momentum for various values of $\alpha$. } 
\end{figure}

\section{spin excitations with the intralayer pairing}

\begin{figure} [htb]
	\centering 
	\includegraphics[width = 10cm]{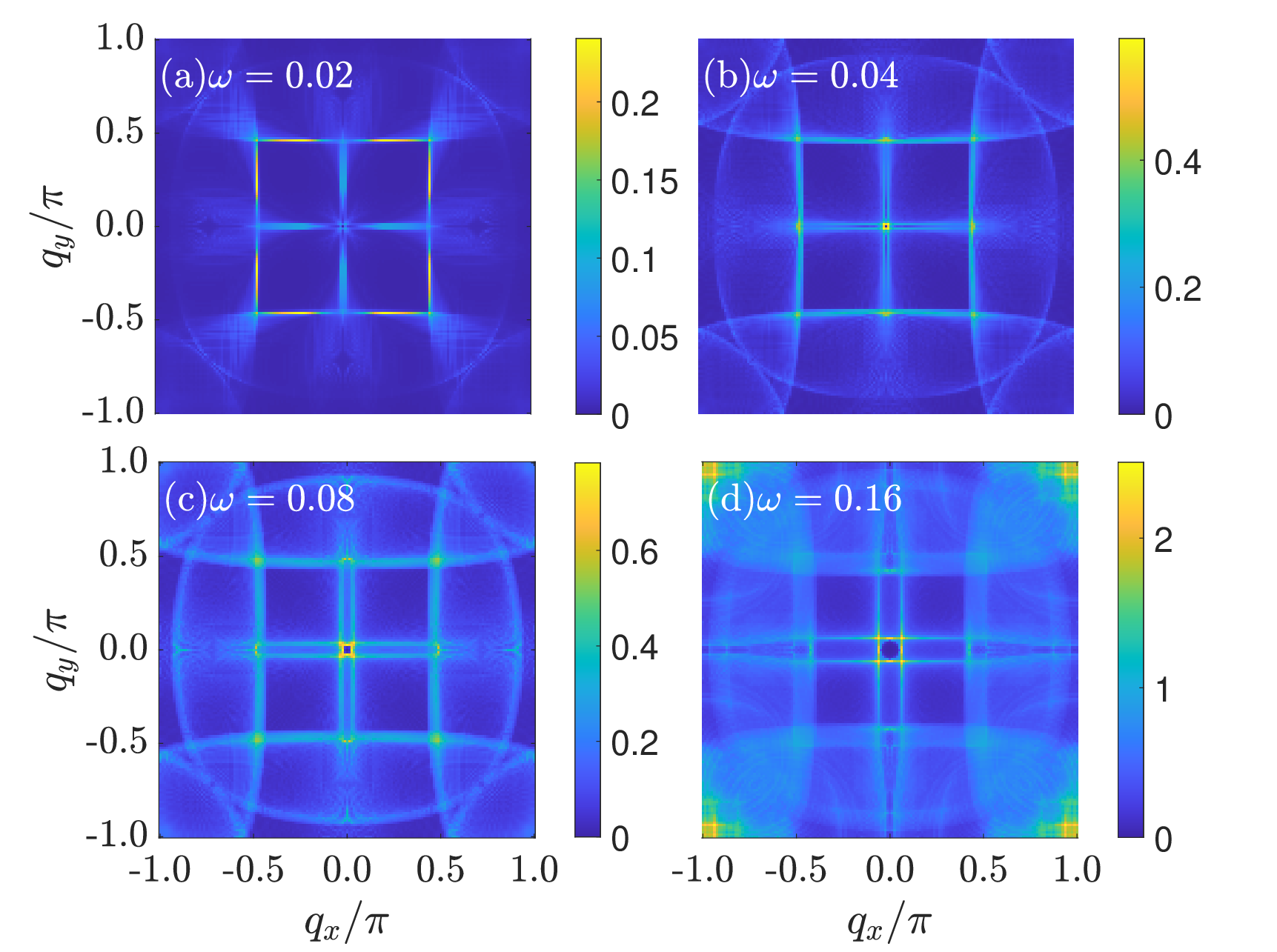} 
	\caption{Imaginary parts of the spin susceptibility with $d$-wave symmetry, plotted as a function of momentum $\mathbf{q}$ for different energies. } 
\end{figure}

\begin{figure} [htb]
	\centering 
	\includegraphics[width = 10cm]{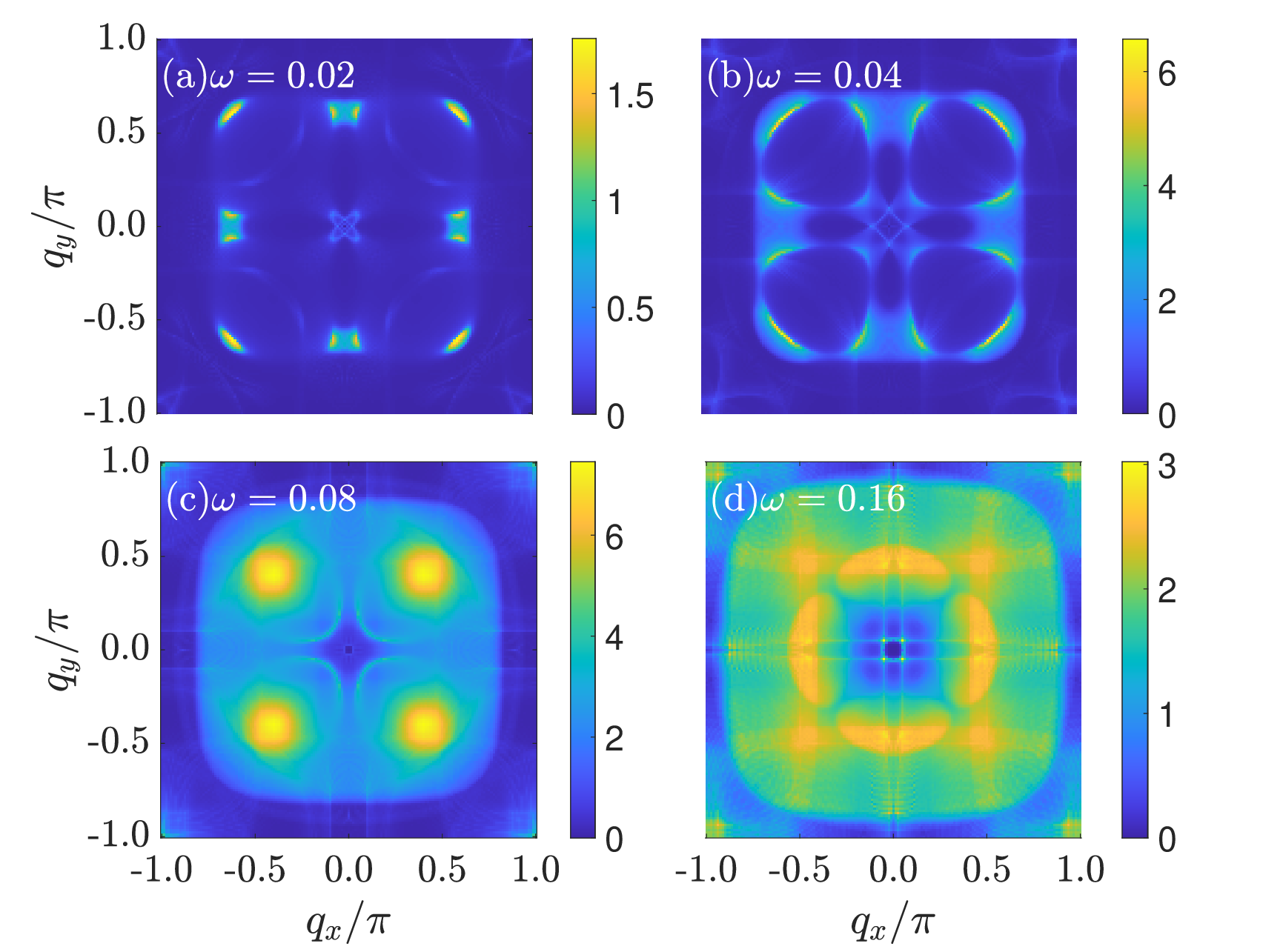} 
	\caption{Imaginary parts of the spin susceptibility with $s_{\pm}$-wave symmetry, plotted as a function of momentum $\mathbf{q}$ for different energies.}
\end{figure}

In the main text, we have presented numerical results for spin excitations with interlayer pairing. Given that dominant intralayer pairing with extended $s_\pm$-wave and $d$-wave pairing symmetries have also been proposed by some theoretical groups, we now investigate the spin excitations for intralayer extended $s$-wave and $d$-wave symmetries.

For the dominant intralayer extended $s$-wave pairing symmetry, we set the intra-layer pairing interactions to $V_{x\parallel} = V_{z\parallel} = 0.6$ and the inter-layer pairing interactions to $V_{x\perp} = V_{z\perp} = 0.3$. The order parameters are then calculated self-consistently using Eqs. (7) and (8) from the main text. The $d$-wave pairing symmetry, which is based on the spin fluctuation scenario as referenced in \cite{PhysRevB.108.L201121}, has been numerically verified not to be supported within the mean-field approach. However, we have obtained the $d$-wave pairing non-selfconsistently, with the gap magnitudes matching those of the $s_{\pm}$ pairing symmetry, as detailed in \cite{PhysRevB.108.174501}.

Subsequently, the intensity plots of the imaginary part of the spin susceptibility as a function of momentum for different energies are presented for both the extended $s$-wave and $d$-wave pairing symmetries in Figs. S2 and S3, respectively.
 
\begin{figure} 
	\centering 
	\includegraphics[width = 14cm]{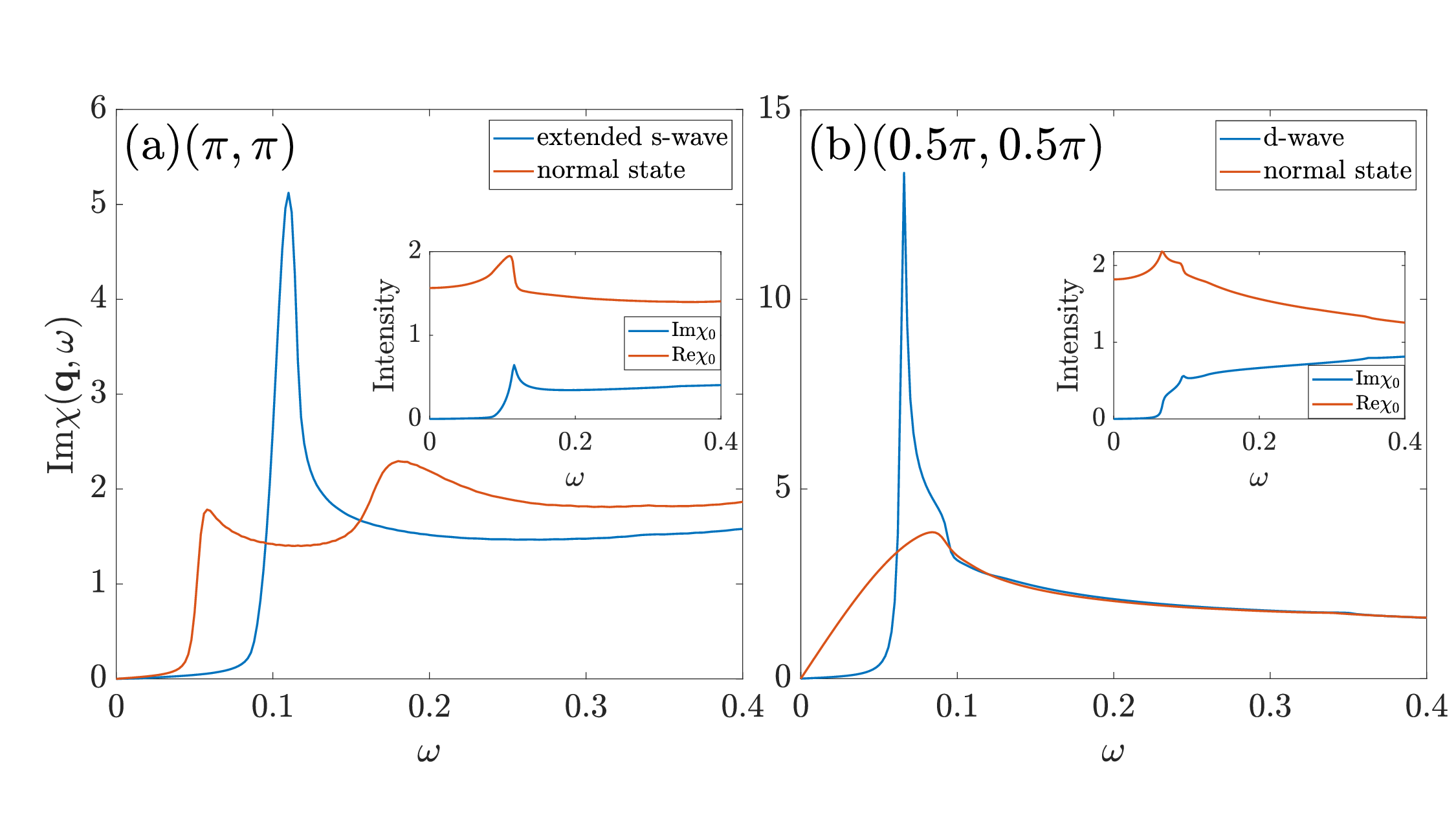} 
	\caption{Imaginary parts of the spin susceptibility plotted as a function of energy, considering different intralayer pairing symmetries. Inset: the real and imaginary parts of the bare spin susceptibility without the RPA correction. }
\end{figure}

For the extended $s$-wave pairing symmetry, at low energies with $\omega = 0.02$, spin excitations are predominantly peaked near the momenta $(\pi/2, \pi/2)$ and $(\pi/2, \pi/4)$. As the energy increases, these excitations shift primarily to $(0, 0)$, $(\pi/2, \pi/2)$, and $(\pi/2, 0)$. Upon further energy increase to 0.16, new spin excitation channels emerge near the momentum $(\pi, \pi)$.

For the $d$-wave pairing symmetry, at low energies with $\omega = 0.02$, spin excitations are mainly observed near $(\pi/2, \pi/2)$ and $(\pi/2, 0)$. As the energy increases, the spin excitations near $(\pi/2, \pi/2)$ significantly enhance. When the energy reaches $\omega = 0.16$, multiple spin excitation channels become evident, primarily near $(\pi/2, \pi/2)$, $(\pi/2, 0)$, and $(\pi, \pi)$. Our findings suggest that at low energies, spin excitations are strongly influenced by the superconducting pairing, which can be instrumental in distinguishing between different pairing symmetries.

The imaginary part of the spin susceptibility as a function of energy at the momentum $\mathbf{q} = (\pi, \pi)$ for the extended $s$-wave superconducting state is depicted in Fig. S4(a). The corresponding results at the momentum $\mathbf{q} = (\pi/2, \pi/2)$ for the $d$-wave pairing state are shown in Fig. S4(b). For comparative analysis, we also present the spin susceptibility in the normal state. It is evident that in the extended $s$-wave superconducting state, a spin resonance peak emerges near the energy $\omega = 0.1$. Similarly, in the $d$-wave superconducting state, a spin resonance peak is observed near the energy $\omega = 0.07$. The underlying mechanism becomes evident when examining the bare spin susceptibility shown in the insets. The characteristic step-like enhancement in Im$\chi_0$ at the spin gap edge originates from the sign-reversed superconducting gaps connected by the nesting vector. Through the Kramers-Kronig relations, this abrupt increase induces a peak structure in the real part Re$\chi_0$, driving the real part of the RPA factor ($\mid \hat{I} -\alpha \hat{U} \mathrm{Re}\hat{\chi}_0(\textbf{q},\omega) \mid$) to approach its minimal value. This critical reduction of the denominator significantly amplifies the spin susceptibility, leading to the observed resonance peaks in the renormalized susceptibility.
 
Our results clearly indicate that spin resonance arises for unconventional off-site intralayer pairing states, with the resonant momentum being dependent on the pairing symmetry. In contrast, for interlayer pairing states, the spin resonance phenomenon is absent. Thus, spin resonance phenomena may serve as a diagnostic tool to identify the pairing symmetry of La$_3$Ni$_2$O$_7$ superconductors.
\end{CJK}
\end{document}